\documentclass[letter,traditabstract]{aa} 
\usepackage{graphicx}
\usepackage{txfonts}
\usepackage{natbib}
\usepackage{threeparttable} 

\begin{document}
\title{
Direct Estimate of Cirrus Noise in \textit{Herschel} Hi-GAL Images
\thanks{\textit{Herschel} is an ESA space observatory with science instruments provided
by European-led Principal Investigator consortia and with important participation from NASA.}
}

\author{
P.~G. Martin\inst{1, 2}
\and
M.-A.~Miville-Desch{\^e}nes\inst{3}
\and
A. Roy\inst{2}
\and
J.-P. Bernard\inst{4}
\and
S. Molinari\inst{5}
\and 
N. Billot\inst{6}
\and 
C. Brunt\inst{7}
\and 
L. Calzoletti\inst{8}
\and 
A.M. DiGiorgio\inst{5}
\and 
D. Elia\inst{5}
\and  
F. Faustini\inst{8}
\and  
G. Joncas\inst{9}
\and  
J.C. Mottram\inst{7}
\and 
P. Natoli\inst{10}
\and 
A. Noriega-Crespo\inst{11}
\and 
R. Paladini\inst{11}
\and  
J.F. Robitaille\inst{9}
\and 
F. Strafella\inst{12}
\and 
A. Traficante\inst{10}
\and
M. Veneziani\inst{13}
}

\institute{
Canadian Institute for Theoretical Astrophysics, University of Toronto, 60 St. George Street, Toronto, ON M5S 3H8, Canada\\
\email{pgmartin@cita.utoronto.ca}
\and
Department of Astronomy \& Astrophysics, University of Toronto, 50 St. George Street, Toronto, ON M5S 3H4, Canada 
\and
Institut d' Astrophysique Spatiale, UMR8617, Universit{\'e} Paris-Sud, F-91405, Orsay, France
\and
Centre d’Etude Spatiale du Rayonnement, CNRS, Toulouse, France
\and
INAF-Istituto Fisica Spazio Interplanetario, Roma, Italy
\and
NASA Herschel Science Center, Caltech, Pasadena, CA
\and
School of Physics, University of Exeter, Stocker Road, Exeter, EX4 4QL, UK
\and
ASI Science Data Center, I-00044 Frascati (Roma), Italy
\and
Departement de Physique, Universit\'e Laval, Qu\'ebec, Canada
\and
Dipartimento di Fisica, Universit\'{a} di Roma 2 ``Tor Vergata'', Roma, Italy
\and
Spitzer Science Center, California Institute of Technology, Pasadena, CA
\and
Dipartimento di Fisica, Universit\'{a} del Salento, Lecce, Italy
\and
Dipartimento di Fisica, Universit\'{a} di Roma 1 ``La Sapienza'', Roma, Italy
}

\date{Received 31 March 2010; accepted 13 May 2010}


\abstract{ In \textit{Herschel} images of the Galactic plane and many star
  forming regions, a major factor limiting our ability to extract faint
  compact sources is cirrus confusion noise, operationally defined as
  the ``statistical error to be expected in photometric measurements due
  to confusion in a background of fluctuating surface brightness.''  The
  histogram of the flux densities of extracted sources shows a
  distinctive faint-end cutoff below which the catalog suffers from
  incompleteness and the flux densities become unreliable.  This
  empirical cutoff should be closely related to the estimated cirrus
  noise and we show that this is the case.  We compute the cirrus noise
  directly, both on \textit{Herschel} images from which the bright sources have
  been removed and on simulated images of cirrus with statistically
  similar fluctuations.  We connect these direct estimates with those
  from power spectrum analysis, which has been used extensively to
  predict the cirrus noise and provides insight into how it depends on
  various statistical properties and photometric operational parameters.
  We report multi-wavelength power spectra of diffuse Galactic dust
  emission from Hi-GAL observations at 70 to 500\,$\mu m$ within
  Galactic plane fields at $l= 30 \degr$ and $l= 59 \degr$.  We find
  that the exponent of the power spectrum is about $-3$.  At
  $250\ \mu{\rm m}$, the amplitude of the power spectrum increases
  roughly as the square of the median brightness of the map and so the
  expected cirrus noise scales linearly with the median brightness.  For
  a given region, the wavelength dependence of the amplitude can be
  described by the square of the spectral energy distribution (SED) of
  the dust emission. Generally, the confusion noise will be a worse
  problem at longer wavelengths, because of the combination of lower
  angular resolution and the rising power spectrum of cirrus toward
  lower spatial frequencies, but the photometric signal to noise will
  also depend on the relative SED of the source compared to the cirrus.}


   \maketitle
%

\section{Introduction}
Cirrus noise, which is operationally defined as the ``statistical error
to be expected in photometric measurements due to confusion in a
background of fluctuating surface brightness'' \citep{gautier}, is a
major issue limiting the cataloging of compact sources that underpins
the study of the early stages of star formation in the interstellar
medium.  Examination of wide range of mass of the stellar precursors,
requires measurement of sources with a wide range of luminosity, or at
each wavelength, flux density. Stars form where there is abundant
material, and so the cirrus brightness in the field tends to be
high. Furthermore, many studies, both targeted and unbiased, are in the
Galactic plane, which is also bright.  Cirrus noise varies with cirrus
brightness (Sect.~\ref{sect:cnps}), introducing further complexity to the
problem.
Cirrus fluctuations characteristically decrease with decreasing spatial
scale (Sect.~\ref{sect:cnps}), but even with the improved angular
resolution
of \textit{Herschel}, 
cirrus noise remains a dominant factor.
The \textit{Herschel} observation planning tool HSpot
(www.ipac.caltech.edu/Herschel/hspot.shtml) has a built-in confusion
noise estimator to provide ``on-line guidance on where to expect
fundamental detection limits for point sources that cannot be improved
by increasing the integration time.'' 
With such guidance the serious impact of cirrus noise was anticipated,
and as shown below it can now be quantified directly using submillimeter
data.

To appreciate the problem at its fundamental level, consider actual
catalogs of extracted sources \citep{MolinariThisVolume,mol2010detect}
in two degree-sized Hi-GAL Galactic plane fields \citep{molinari2010}
(see Sect.~\ref{sect:obs}).
\begin{figure}
    \centering
      \includegraphics[width=0.4\textwidth, angle=0]{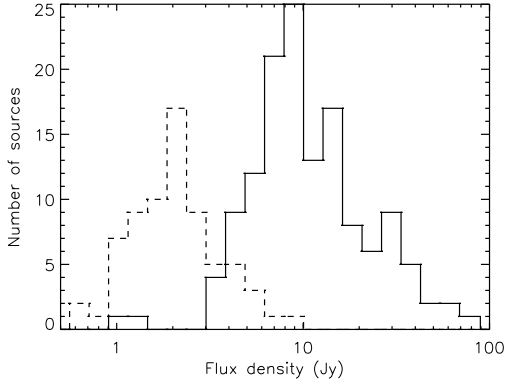}
      \caption{ Solid histogram: 136 sources cataloged at 250\,$\mu m$
        in a representative large degree-sized sub-region of Hi-GAL $l=
        30\degr$ field (f30: see Sect.~\ref{sect:obs}).  Dashed histogram:
        70 sources cataloged in a sub-region of Hi-GAL $l= 59\degr$
        field (f59) that has a factor 5.8 lower median brightness.}
          \label{fig:hist2}
\end{figure}
The solid histogram in Figure~\ref{fig:hist2} shows the flux densities
of sources cataloged at 250\,$\mu m$ in the brighter field.  Note the
falloff in source counts at flux densities less than 10~Jy. This falloff
is not an intrinsic property of the underlying population of sources.
We show below that this falloff is the expected consequence of cirrus
noise, which can be quantified independently of the making of the
catalogs.  In the fainter field (dashed histogram), the cutoff is lower
because of decreased cirrus noise.
Clearly, it will be important to account for the varying dramatic
effects of cirrus noise in order to recover the statistics of the
intrinsic faint-source population.

Cirrus noise is operationally defined and so for a particular measuring
strategy, such as fitting compact sources with Gaussians as used in
Hi-GAL, this can actually be estimated
directly from the source-removed maps (Sect.~\ref{sect:cnps}).
There is also an extensive literature on ``estimating confusion noise
due to extended structures given some statistical properties of the
sky'' \citep{gautier}; see
\citep{kiss-galcir,jeong2005,ma-dust,roy2010}.  The cirrus brightness
statistics are well described by a power spectrum, appropriate for
Gaussian random fields.  \citet{gautier} found some non-Gaussianity and
\citet{ma-dust} found non-vanishing skewness and excess kurtosis in the
underlying brightness fluctuation fields. Nevertheless, these are not
large effects and for estimating the variance the power spectrum is
demonstrably still a powerful tool, particularly because of the insight
it provides into how the cirrus noise depends on various statistical and
operational parameters.
Therefore, we connect the new direct estimates with what can be obtained
using power spectra (Sect.~\ref{sect:cnps}).  In Sect.~\ref{sect:prop} we show
that reliable power spectra can be obtained even from first-generation
images processed for Hi-GAL.
Finally, we return to how estimates of cirrus noise should ultimately
explain the faint-end cutoff in forthcoming source catalogs
(Sect.~\ref{sect:noise}).

\section{Observations}\label{sect:obs}

Two fields were observed (at $l\sim30\degr$ and $l\sim59\degr$) during
SDP as part of Hi-GAL \citep{molinari2010}, using the Parallel mode with
both nominal and orthogonal scanning to acquire data in the PACS\,70
and 160\,$\mu m$ bands and all three SPIRE bands (250, 350 and
500\,$\mu m$).
The data processing using HIPE and map-making using ROMAGAL are
described by \citet{MolinariThisVolume} and references therein.
The maps were converted into brightness units (MJy/sr) and the DC
offsets recovered as described by \citet{BernardThisVolume}.
The offsets are not needed for the cirrus noise analysis, but are needed
to describe, for example, the median brightness of the map used in
Sect.~\ref{sect:cnps} and Sect.~\ref{sect:prop}.

Compact sources have been detected and then quantified using a
Gaussian model \citep{MolinariThisVolume,mol2010detect}. Using the modeled
properties, the sources have been removed to produce the images used
for the analysis here.
There remain some bright, fairly compact peaks not meeting the
classification criteria for compact sources, but probably
gravitationally influenced and not standard cirrus structure.  There are
also some artefacts from removal of the brightest compact sources.
Given the preliminary nature of these source-subtracted images, for this
initial analysis we have analysed large degree-sized sub-regions in
which these effects are minimized.  They are otherwise representative of
bright cirrus.
The brightest is from the fairly homogeneous low-longitude half of the
$l\sim30\degr$ map, delineated by $29 \la l \la 29.83 \degr$ and $-0.5
\la b \la 0.33\degr$, with median brightness $<I_{250}>\ = 1390\ {\rm
  MJy}\ {\rm sr^{-1}}$.  For convenience, we refer to this field as
f30.
A contrasting region, in having a lower median brightness $241\ {\rm
  MJy}\ {\rm sr^{-1}}$, is the lower right quarter $l\sim59\degr$ map,
delineated by $58 \la l \la 59 \degr$ and $-1 \la b \la 0\degr$ (denoted
f59).

\section{Direct estimates of cirrus noise and relationship to power spectra}\label{sect:cnps}

In the literature, cirrus noise is defined operationally for a
photometric measurement template used to evaluate the flux density of a
compact source on a fluctuating background (not simply ``finding'' it).
We obtain a direct estimate of cirrus noise numerically by placing the
template randomly in a cirrus map and finding the rms of the apparent
``source'' flux density, $\sigma_{\rm{cirrus}}$. 
In the context of power-law cirrus, \citet{gautier} quantified this
analytically for several measurement templates, and these plus simulated
cirrus maps have been used to validate our numerical approach in detail
(see below).  Among these templates is the ``aperture plus reference
annulus'' chosen by \citet{helou1990} and adopted by \citet{kiss2001}
and \citet{roy2010}.
In practice, compact sources are often fit with a model template
consisting of a Gaussian and a planar inclined plateau, with a footprint
larger than the extent of the Gaussian to provide a reference area to
characterize the ``background.'' We adopt a radius of 1.82 times the
FWHM of the Gaussian used.  This Gaussian-based strategy has the
practical advantage of still being useful for moderately crowded
sources, unlike simple aperture plus annulus photometry.
As a specific and relevant example using the 250\,$\mu m$ Hi-GAL fields,
the source FWHM is typically 1.5 times larger than the
diffraction-limited PSF and so we adopt a Gaussian template with this
larger FWHM.  We also assume that the source-subtracted map is a good
proxy for the cirrus in the field.  The estimated cirrus noise is 1.7~Jy
for f30 and 0.23~Jy for the fainter f59, scaling roughly as the median
brightness which changes by a factor 5.8.

As mentioned, there is extensive literature in which the cirrus noise is
quantified using the power spectrum. Furthermore, these authors showed
that the power spectrum of Galactic cirrus follows a power law
$P_{\rm{cirrus}}(k)=P(k_0)(k/k_0)^{\alpha}$, quantified by an amplitude
$P_0\equiv P(k_0)$ at some fiducial $k_0$, and an exponent $\alpha$ that
is typically $-3$ \citep{ma-dust}, as also found here.  We adopt $k_0 =
1.0\ {\rm arcmin}^{-1}$, a scale close to the beam sizes, to avoid
issues of extrapolation if $\alpha$ is not quite $-3$.
The essentials of the detailed \citet{gautier} analysis of the ``cirrus
noise'' for a telescope with mirror diameter $D$ working at wavelength
$\lambda$ and $\alpha = -3$ can be summarized as:
\begin{eqnarray}
\sigma_{\rm{cirrus}} = A_t\ R_t^{2.5}
\left(\frac{\lambda/250\ \mu{\rm m}}{D/3.5\ {\rm m}}\right)^{2.5}
\left(\frac{P_0}{10^{6}\ {\rm Jy}^2\ {\rm sr}^{-1}}\right)^{0.5}\ {\rm Jy}.
\label{eq:sigcir}
\end{eqnarray}
Here $R_t$ is a dimension of the measurement template (e.g., aperture
diameter or FWHM of Gaussian) in units of $\lambda/D$ (explicitly in
the formula with the same power).  If the compact sources being measured
are extended, as will be the case in many Galactic surveys including
those illustrated here, then $R_t$ needs to be correspondingly larger.
The consequent increase in $\sigma_{\rm{cirrus}}$ is captured by the
factor $R_t^{2.5}$ for the range of interest.
$A_t$ is an amplitude appropriate to a particular measurement template.
$A_t$ actually changes significantly with $R_t$ for small $R_t$, but
this is not relevant because no practical application would use
photometric templates smaller than the diffraction limit; for the range
of interest, there is only a weak $R_t^{0.3}$ dependence (see also
Fig.~3 in \citealp{gautier}).
Neither $A_t$ or $R_t$ depend on wavelength; that dependence is carried
explicitly in the $\lambda/D$ term and implicitly in the square root
dependence on $P_0$.

We have made simulated power-law cirrus maps with $\alpha = -3$ and
directly evaluated the cirrus noise with various templates for a range
of $R_t$. We recover the predicted $R_t^{2.5}$ dependence and specific
values of $A_t$, like 0.034 predicted for the widely-used
aperture-annulus template of \citet{helou1990} with $R_t$ near 1.6.
From these simulated maps we also derived the amplitude appropriate to
the Gaussian fitting template for the empirical range of $R_t$ near 1.8
for the actual sources. When the simulated cirrus map has resulted from
convolved with the actual Neptune beam (Sect.~\ref{sect:prop}), $A_t =
0.054$ ($A_t = 0.065$ for the nominal Gaussian PSF).

\section{Properties of observed power spectra}\label{sect:prop}

Details of computing the power spectrum and its errors may be found, for
example, in \citet{roy2010}.
In practice, contributions to the total power spectrum come not only
from diffuse dust emission, but also from point sources, the cosmic
infrared background (CIB), and ``noise'' (which might reasonably be
assumed to be fairly white).  When these components are statistically
uncorrelated, the total power spectrum can be expressed as
\citep{ma-dust}:
\begin{equation}
P(k)=\Gamma(k)\left[P_{\rm{cirrus}}(k)+P_{\rm{source}}(k)+P_{\rm{CIB}}(k)\right]+N(k).
 \label{eq:power}
\end{equation}
For bright Galactic plane fields, the contribution from the CIB to the
power spectrum is negligible. We have removed the brighter compact
sources, but others at or below the detection threshold must remain and
need to be accounted for by $P_{\rm{source}}$.  While the ``noise'' $N$
is measurable, and of interest, it too makes an insignificant
contribution.

$\Gamma(k)$, the power spectrum of the PSF, decays at large $k$.  For
the SPIRE bands we used the empirical PSF from scans of Neptune
(ftp://ftp.sciops.esa.int/pub/hsc-calibration/SPIRE/PHOT/Beams) and
computed $\Gamma(k)$ using the technique developed by \citet{roy2010}. A
Gaussian approximation is poor and adversely impacts the extraction of
$P_{\rm{cirrus}}(k)$, but the fitted functions in
Table~\ref{table:gamma} are good to within a few percent for the range
of interest $k \ < k_{max}/2$, where $ k_{max}$ is the highest spatial
frequency in the power spectrum for these approximately Nyquist-sampled
maps ($5\ {\rm arcmin}^{-1}$ at 250\,$\mu m$).

\begin{table}
\caption{Parameters$^a$ for SPIRE $\Gamma(k)$ at 250, 350, 500\,$\mu m$}            
\label{table:gamma}      
\centering          
  \begin{threeparttable}
\begin{tabular}{l l l l l}
\hline\hline
$a_0$ &$a_1$ &  $a_2$ & $a_3$ & $a_4$ \\ 
\hline
0.789137 & $-0.466966$ & -0.070793 & 0.349304 & $-0.068133$ \\
0.671082 & $-0.807737$ &  0.294573 & 0.250785 & $-0.125454$ \\
0.402564 & $-0.897627$ &  0.095114 & 1.785801 & $-0.720592$ \\
\hline          
\end{tabular}
    \begin{tablenotes}
       \item[a] $\Gamma(k) = [1 + \sum_{i=1,4} a_i k^i ] \times \exp[-k^2/(2 a_0^2)]$ 
     \end{tablenotes}
  \end{threeparttable}
\end{table}

\begin{figure}
    \centering
      \includegraphics[width=0.45\textwidth, angle=0]{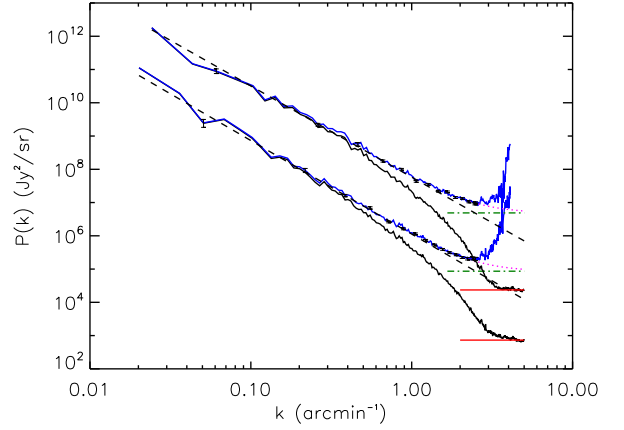}
      \caption{
Top curves: power spectrum at 250\,$\mu m$ for the f30 field
(Sect.~\ref{sect:obs}).
Decay at large $k$ in lower of the two curves, $P(k)$, is due to
$\Gamma(k)$ from the PSF.
Plateau at high $k$ is from residual ``noise'' $N$ in the map,
$2.4\ \times 10^4\ {\rm Jy}^2\ {\rm sr^{-1}}$.
At very small scales, where the astronomical signals become correlated
within the beam, $P$ meets $N$.
Upper curve is $P_{\rm{cirrus}}(k) + P_{\rm{source}}$, after subtracting
the noise and dividing by $\Gamma(k)$ to remove the effect of the beam.
Upper curve has been fit over the range $0.06\ {\rm{arcmin^{-1}}} <$ $k$ $ <
2.5\ {\rm{arcmin^{-1}}}$ by a simple model (dotted curve) consisting of
a power law (dashed line) with $P_0\ =\ (58 \pm 1)\ \times 10^6\ {\rm
  Jy}^2\ {\rm sr^{-1}}$ plus a constant $P_{\rm{source}}\ =\ (4.9 \pm
0.3) \ \times 10^6\ {\rm Jy}^2\ {\rm sr^{-1}}$ (dash dot line).
The power-law exponent is $-2.74 \pm 0.03$.
%
%
Bottom curves: power spectrum for the fainter f59 field.
The ``noise'' level, $P_0$, and $P_{\rm{source}}$ have all decreased,
roughly as the square of the median brightness, to $7.3\ \times 10^2$ and
$(1.13 \pm 0.03)\ \times 10^6$ and $(8.6 \pm 0.6)\ \times 10^4\ {\rm Jy}^2\ {\rm
  sr^{-1}}$, respectively.
The power-law exponent $-2.81 \pm 0.03$ is not significantly different.
}
\label{fig:powspec}
\end{figure}

The top set of curves in Figure~\ref{fig:powspec} shows the power
spectrum at 250\,$\mu m$ for the f30 field.
Compared to the power spectra from BLAST shown by \citet{roy2010},
$P(k)$ can be determined to larger $k$, as expected because of the smaller
\textit{Herschel} beam.
The lower set of curves shows similar results
for the fainter f59 field.
The steps in fitting the data by equation~\ref{eq:power} to find
$P_{\rm{cirrus}}(k)$, and the results of the fit, are described in the
figure caption.  The fit is very good (reduced $\chi^2 \sim 1.5$).

For the fainter f59 field, the noise level, $7.3\ \times 10^2\ {\rm
  Jy}^2\ {\rm sr^{-1}}$, is comparable to the detector noise predictable
by HSpot. It is in fact slightly lower than for the much fainter Polaris
field observed with the same coverage (see Fig.~3 in
\citealp{MamdThisVolume}).  Tests indicate that this slight improvement
comes from using ROMAGAL rather than the naive map-maker in HIPE.
However, note that the noise level for the brighter sub-region is
dramatically higher, $2.4\ \times 10^4\ {\rm Jy}^2\ {\rm sr^{-1}}$.  This
appears to scale as the square of the median brightness of the map and
so must measure residual map artefacts, not detector noise.
Nevertheless, these are already excellent images with the ``noise''
small compared to $P_{\rm{cirrus}}(k)$ over the range of interest $k\ <
2.5\ {\rm{arcmin^{-1}}}$.  The presence of $P_{\rm{source}}$ is only
subtly evident in a slight concave curvature of the power spectrum at
large $k$ and is adequately modeled there as a constant.

For f30 we find $\alpha = -2.74 \pm 0.03$ and in the fainter f59 field
$-2.81 \pm 0.03$, not significantly different.  These are close to
estimates in the Galactic plane by \citet{roy2010}, and also remarkably
similar to what has been found at 100\,$\mu m$ for high-latitude diffuse
cirrus \citep{ma-dust}, despite quite different conditions and
geometries which could in principle affect the cirrus fluctuations.
This supports the adoption of $\alpha = -3$ for equation~\ref{eq:sigcir}
to quantify $\sigma_{\rm{cirrus}}$ in the submillimeter.

At 250\,$\mu m$ we have examined two further regions, the upper right
quarter of the $l= 59\degr$ map and the Polaris field.  From this we
conclude that $P_0$ scales \textit{roughly} as the square of the median
brightness.  A corollary is that $\sigma_{\rm{cirrus}}$ should scale
roughly as the median brightness, as we have already found directly
(Sect.~\ref{sect:cnps}).  This behavior is as though the inherent
statistical structure in brightness $I_\nu$ is the same, just scaled up
and down.  Empirically, at 100\,$\mu m$ a steeper scaling has been seen
for bright cirrus \citep{ma-dust}, and so this should be revisited for
all wavelengths when more fields are available.  One factor that is
changing from field to field, and within fields, is the dust temperature
$T_{\rm{d}}$ \citep{BernardThisVolume}.  Even with the same dust column
density structure, this will modulate the cirrus brightness, with
stronger effects at 100\,$\mu m$ than in the submillimeter.

For a given region, images of optically thin dust emission at different
passbands could be simply related by the scale factor $\kappa_\nu
B_\nu(T_{\rm{d}})$ or the (relative) spectral energy distribution (SED).
Here, $\kappa_\nu$ is the dust emissivity.  Because of averaging along
the line of sight and over different grain components with potentially
different $T_{\rm{d}}$, this is a simplification. Nevertheless,
\citet{roy2010} demonstrated with BLAST and IRAS data how an SED with a
reasonable $T_{\rm{d}}$ can be recovered from the wavelength dependence
of $P_0^{1/2}$ (conversely, $P_0$ varies as SED$^2$).

\begin{figure}
\includegraphics[scale=0.3,angle=0]{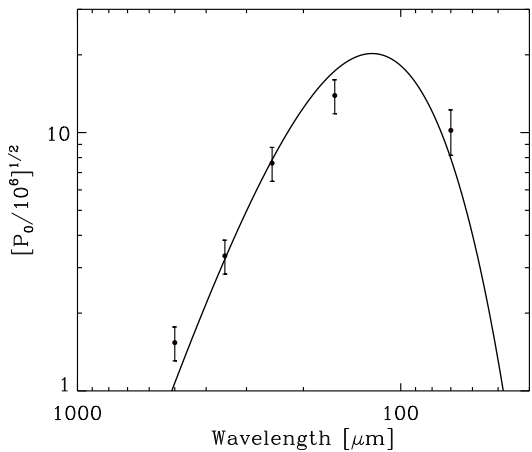}
\caption{SED obtained from the square root of the amplitudes $P_0$
  obtained from fits to power spectra in the f30 field.
  Solid curve is from the fit of a $\beta\ =\ 2$ modified blackbody,
  with temperature $23.6$~K.}
\label{fig:psed}
\end{figure}

We have examined this with the \textit{Herschel} data, measuring the power
spectra and finding $P_0$ for each of the five passbands.  The results
for the $l\sim30\degr$ sub-region are plotted in Figure~\ref{fig:psed}.
These data were then fit with a single-temperature modified blackbody
($\kappa_\nu \propto \nu^2$ or $\beta = 2$).  We find that the
functional form is satisfactory, and that $T_{\rm{d}} = 23.6 \pm 1.0$~K.
The temperature has been fit independently pixel by pixel
\citep{BernardThisVolume}.  For this sub-region, with $\beta = 2$ the
average temperature is 22.5~K (range 16 -- 28.5~K, median: 22.5~K).  The
close agreement is reassuring.  Thus, other things being equal, the SED
of $\sigma_{\rm{cirrus}}$ will be close to that of the cirrus
emission. In measuring the flux density of a compact source, the
wavelength dependence of the signal (source SED) to noise (diffuse dust
SED) will depend on the relative temperatures as well.  Sources cooler
(hotter) than the cirrus will be detected at a relatively higher (lower)
S/N at long wavelengths.  However, equation~\ref{eq:sigcir} also shows
that in practice cirrus noise will be more severe for the
long-wavelength bands because of the larger beam, offsetting this effect
on S/N for cooler sources and compounding it for hotter ones.

\section{Cirrus noise and faint-end cutoffs in catalogs}\label{sect:noise}

We can estimate the cirrus noise from the power spectra and
equation~\ref{eq:sigcir}. For f30 and f59 at $P_0$ is 58 and
$1.1\ \times 10^6\ {\rm Jy}^2\ {\rm sr^{-1}}$, respectively.  When fit by
Gaussians, sources in these Hi-GAL fields require typically $R_t = 1.8$.
$A_t = 0.054$ has been evaluated from simulations for this range of
$R_t$.  We find $\sigma_{\rm{cirrus}} = 1.8$ and 0.25~Jy, respectively.
These compare favorably with the values estimated directly, 1.7 and
0.23~Jy, respectively (Sect.~\ref{sect:cnps}).
In the f30 field, a 5-$\sigma$ catalog would be cut off at about 9~Jy.
The fainter f59 field is also affected, but would have a lower
5-$\sigma$ cutoff of 1.3~Jy.  Cirrus noise therefore severely limits the
depth of the catalogs and generally in Hi-GAL Galactic plane fields will
be the pre-dominant factor.
When catalogs from multiple fields with different median brightnesses
are combined/interpreted, account must be made for the differing impact
of cirrus noise.

The faint-end cutoffs in the catalogs of compact sources illustrated in
Figure~\ref{fig:hist2} are close to the estimated influence of cirrus
noise.  Quantitatively, this must be somewhat of a coincidence. As
described by \citet{MolinariThisVolume} and \citet{mol2010detect}, the
catalog is not selected on the basis of signal to noise (S/N) of the
flux densities.  No errors have been tabulated for the flux densities
from the Gaussian fits. 
Judging from their \textit{a posteriori} S/N estimator, which involves
peak flux (Jy/beam), there is a wide range, including some below 5.
Simulations to assess catalog completeness find that the peak flux for
80\% completeness is 4.1~Jy/beam over the entire $l\sim30\degr$ map, and
so for the typical source FWHM, the corresponding flux density
completeness limit is 9.2~Jy close to our estimated 5-$\sigma$ threshold
and the observed histogram peak in Figure~\ref{fig:hist2}.  For the
fainter $l\sim59\degr$ survey there is a lower 80\% completeness limit
(1.6~Jy); peak in the histogram occurs at a lower value, again close to
this limit and our estimated 5-$\sigma$ threshold.

At longer wavelengths, cirrus noise will be more limiting for both
source detection and source flux density determinations, because of the
$\lambda^{2.5}$ (beam) dependence.  This highlights another important
consideration for band-merged catalogs, that the S/N for cataloged
sources will be wavelength dependent.



\end{document}